\documentclass[runningheads]{elsarticle}

\usepackage{graphicx}
\usepackage{amsmath, amssymb, amsthm}
\usepackage{lineno}

\newtheorem{definition}{Definition}

\newtheorem{theorem}{Theorem}

\newcommand{\rationals}{\mathbb{Q}}
\newcommand{\naturals}{\mathbb{N}}

\newcommand{\diag}{\text{diag}}

\newcommand{\seq}[1]{\langle#1\rangle}

\begin{document}

\begin{frontmatter}
\title{Skolem and Positivity Completeness of Ergodic Markov Chains}

\author{Mihir Vahanwala\fnref{fn1}\affiliation{organization={Max Planck Institute for Software Systems, Saarland Informatics Campus},
            country={Germany}}}
\fntext[fn1]{The author is partially funded by DFG grant 389792660 as part of TRR 248 (see https://perspicuous-computing.science).}

\begin{abstract}
We consider the following Markov Reachability decision problems that view Markov Chains as Linear Dynamical Systems: given a finite, rational Markov Chain, source and target states, and a rational threshold, does the probability of reaching the target from the source at the $n^{th}$ step: (i) equal the threshold for some $n$? (ii) cross the threshold for some $n$? (iii) cross the threshold for infinitely many $n$? These problems are respectively known to be equivalent to the Skolem, Positivity, and Ultimate Positivity problems for Linear Recurrence Sequences (LRS), number-theoretic problems whose decidability has been open for decades. We present an elementary reduction from LRS Problems to Markov Reachability Problems that improves the state of the art as follows. (a) We map LRS to \emph{ergodic} (irreducible and aperiodic) Markov Chains that are ubiquitous, not least by virtue of their spectral structure, and (b) our reduction maps LRS of order $k$ to Markov Chains of order $k+1$: a substantial improvement over the previous reduction that mapped LRS of order $k$ to reducible and periodic Markov chains of order $4k+5$. This contribution is significant in view of the fact that the number-theoretic hardness of verifying Linear Dynamical Systems can often be mitigated by spectral assumptions and restrictions on order.
\end{abstract}


\end{frontmatter}

\section{Introduction}
Markov Chains, owing to their formulation in terms of stochastic matrices, are syntactically a special class of Linear Dynamical Systems. However, it is their semantics as transformers of probability distributions that earns them widespread attention as a natural mathematical framework to describe probabilistic systems. Markov Decision Processes, for instance, extend Markov Chains and are a probabilistic decision-making model fundamental to verification \cite[Chapter 10]{Baier-Katoen} and Reinforcement Learning \cite[Section 3]{mdp-rl}. 

Given their importance, there is naturally an extensive body of work on model checking Markov Chains: see \cite{Baier-Katoen} for a comprehensive set of references. Most of the focus has been on the verification of linear- and branching-time properties of Markov Chains through solving systems of linear equations, or linear programs. An alternative approach \cite{Agrawal2015,logicprob,KVAK10,KwonGul} is to consider specifications on the state distribution at each time step, e.g., whether the probability of being in a given state at the $n^{th}$ step is at least $1/4$. Decidability in this setting is a lot more inaccessible: \cite{Agrawal2015,logicprob} only present incomplete or approximate verification procedures, while \cite{KVAK10,KwonGul} owe their model-checking procedures to additional mathematical assumptions. The following decision problems are representative of the fundamental difficulties this approach is fraught with:

\begin{definition}[Markov Reachability Problems]
Given a stochastic matrix $\mathbf{M} \in \rationals^{k \times k}$ (i.e.\ an order $k$ Markov Chain), a threshold $r \in \rationals$, and indices $i, j$, decide whether: 
\begin{enumerate}
\item There exists an $n \in \naturals$ such that $m_{ij}^{(n)} = r$.
\item There exists an $n \in \naturals$ such that $m_{ij}^{(n)} \sim r$.
\item There exist infinitely many $n \in \naturals$ such that $m_{ij}^{(n)} \sim r$.\footnote{For technical reasons discussed in \S\ref{prelimLRS} of the Preliminaries, deciding whether there exist infinitely many $n \in \naturals$ such that $m_{ij}^{(n)} = r$ is actually a tractable problem.}
\end{enumerate}
where $\sim$ denotes one of $\{>, <\}$ and $m_{ij}^{(n)}$ denotes the entry in the $i^{th}$ row and $j^{th}$ column of $\mathbf{M}^n$.
\end{definition}

It is here that the syntactic nature of Markov Chains as Linear Dynamical Systems is brought to the fore. The essential tool we use to argue the difficulty of these problems is the Linear Recurrence Sequence (LRS). An LRS over $\rationals$ of order $k$ is an infinite sequence $\seq{u_n}_{n=0}^\infty$ of rational numbers satisfying a recurrence relation $$u_{n+k} = a_{k-1}u_{n+k-1}+\dots+a_0u_n$$ for all $n \in \naturals$, where $a_0 \ne 0, a_1, \dots, a_{k-1} \in \rationals$. An LRS over $\rationals$ of order $k$ is uniquely specified by $2k$ rational numbers: $a_0, \dots, a_{k-1}$ for the recurrence relation, and $u_0, \dots, u_{k-1}$ for the initial terms. An LRS can be computed through its companion matrix $\mathbf{A}$:
$$
\mathbf{A}^n\begin{bmatrix}
u_0 \\ u_1 \\ \vdots \\ u_{k-2} \\ u_{k-1}
\end{bmatrix} =
\begin{bmatrix}
0 & 1 & 0 & \dots & 0 \\
0 & 0 & 1 & \dots & 0 \\
\vdots & \vdots & \vdots & \ddots & \vdots \\
0 & 0 & 0 & \dots & 1 \\
a_0 & a_1 & a_2 & \dots & a_{k-1}
\end{bmatrix}^n
\begin{bmatrix}
u_0 \\ u_1 \\ \vdots \\ u_{k-2} \\ u_{k-1}
\end{bmatrix} =
\begin{bmatrix}
u_n \\ u_{n+1} \\ \vdots \\ u_{n+k-2} \\ u_{n+k-1}
\end{bmatrix}.
$$
The behaviour of an LRS is governed by the eigenvalues of $\mathbf{A}$, or equivalently, the roots of its characteristic polynomial $X^k - a_{k-1}X^{k-1} - \dots - a_0$. The LRS is said to be \emph{non-degenerate} if the ratio of two distinct characteristic roots is never a root of unity, and \emph{simple} if the characteristic polynomial has no repeated roots.

The following are key number-theoretic open decision problems for LRS:
\begin{enumerate}
\item Skolem Problem: Does there exist $n \in \naturals$ such that $u_n = 0$?
\item Positivity Problem: Is $u_n \ge 0$ for all $n \in \naturals$?
\item Ultimate Positivity Problem: Is $u_n \ge 0$ for all but finitely many $n \in \naturals$?
\end{enumerate}

Number-theoretic hardness is formally shown for the Positivity and Ultimate Positivity Problems of order $6$ LRS in \cite{joeljames3}: decision procedures would enable one to quantify how well the continued fraction expansion of a given irrational number converges, an endeavour inaccessible to the means of contemporary number theory.

On the other hand, decidability results have been obtained through suitable restrictions. The Skolem Problem is known to be decidable for LRS of order up to $4$, see \cite{skolem,skolemv}. Very recently, there have been conditional decidability results for LRS of order $5$, and simple LRS \cite{bilu2022skolem}. The Positivity Problem and Ultimate Positivity Problems are decidable up to order $5$ \cite{joeljames3}. If we restrict ourselves to simple LRS, then Ultimate Positivity is decidable \cite{ouaknine2014ultimate}, and Positivity is decidable up to order $9$ \cite{ouaknine2014positivity}. Neither decidability nor hardness results are known for the Positivity Problem for simple LRS of order $10$ and above.

Returning to our stochastic setting, we note that techniques to reduce the above LRS problems to Markov Reachability Problems were known since the work of Turakainen \cite{turakainen}, albeit in the context of probabilistic automata\footnotemark.\ The folkloric results are sharpened in \cite{oldresult}: it is shown that the LRS Problems for order $k$ LRS reduce to Markov Reachability Problems for \textit{reducible and periodic} Markov Chains of order $4k+5$. 

\footnotetext{It is not difficult to use the Cayley-Hamilton Theorem to show the reverse reduction, i.e.\ that the Markov Reachability problems reduce to the corresponding LRS Problems.}

Subsequent works on probabilistic systems such as \cite{akshaycav2023, piribauer} that faced the Markov Reachability Problems use \cite{oldresult} or similar arguments to justify the development of solutions that are useful despite being restricted or incomplete, rather than tackling the main problem head on. One can argue that this justification, as it stands, has the following deficiencies that need to be addressed:

\begin{enumerate}
\item The best known reduction demonstrates the hardness of verifying Markov Chains that are reducible and periodic, rather than \textit{ergodic} (irreducible and aperiodic) Markov Chains that the community most often considers.
\item Breakthroughs in the verification of Linear Dynamical Systems often result from restrictions on the dimensions \cite{joeljames3,skolem,skolemv,ouaknine2014positivity} or spectrum \cite{ouaknine2014ultimate,ouaknine2014positivity,prefixindependent,powerpositivity} of the underlying matrix. A strong reduction needs to show that hardness persists despite the additional spectral properties (see, e.g.\ Theorem \ref{spectral}) enjoyed by ergodic Markov Chains, and also be faithful with respect to order while mapping LRS to stochastic matrices.
\end{enumerate}

Our main result comprehensively addresses this issue.

\begin{theorem}[Main Result]
\label{mainresult}
The Skolem, Positivity, and Ultimate Positivity Problems for LRS of order $k$ reduce to Markov Reachability Problems for ergodic Markov Chains of order $k+1$.
\end{theorem}

\section{Preliminaries}
\subsection{Linear Recurrence Sequences}
\label{prelimLRS}

The celebrated Skolem-Mahler-Lech Theorem \cite{skolem-sml, mahler-sml, lech-sml} states that the set $\{n: u_n = 0\}$ is the union of a finite set $Z$ and finitely many arithmetic progressions $c_1 + b_1\naturals, \dots, c_d + b_d\naturals$, where $0 \le c_i < b_i$ for all $i$. The values of $d, c_i, b_i$ can all be effectively computed, whereas determining whether $Z$ is nonempty is precisely the difficulty of the Skolem Problem. It is well known \cite{berstelmignotte,everest} that for any LRS $\seq{u_n}_{n=0}^\infty$, one can compute $B \ge 1$ such that for all $0 \le c < B$, the subsequence $\seq{u_{nB + c}}_{n=0}^\infty$ is a non-degenerate LRS which, by an observation of Lech \cite{lech-sml}, is either identically zero or has finitely many zeroes. {It thus suffices to study \textbf{non-degenerate} LRS to solve the Skolem, Positivity and Ultimate Positivity Problems.}

\subsection{Stochastic Matrices and Markov Chains}
When the dimensions are clear from context, we use $\mathbf{1}$ to denote the column vector whose entries are all $1$, $\mathbf{I}$ to denote the identity matrix, $\mathbf{0}$ to denote the zero column vector, and $\mathbf{O}$ to denote the zero matrix. Superscript $T$ denotes transposition. We use $\mathbf{e_i}$ to denote the elementary column vector, i.e.\ the vector whose $i^{th}$ entry is $1$ and all other entries are $0$, e.g.\ $\mathbf{e_1} = \begin{bmatrix}1 & 0 & \dots & 0\end{bmatrix}^T$. We use $m_{ij}^{(n)}$ as shorthand to denote the entry in the $i^{th}$ row and $j^{th}$ column of the matrix $\mathbf{M}^n$, i.e.\ $m_{ij}^{(n)}  = \mathbf{e_i}^T \mathbf{M}^n \mathbf{e_j}$. When not specified, $n = 1$.

Distributions with finite support are represented by column vectors. Since they denote probabilities of mutually exclusive and exhaustive events, their entries are non-negative and sum up to $1$. A finite $k$-state Markov Chain is given by a (left) stochastic matrix $\mathbf{M} \in \rationals^{k \times k}$, i.e.\ all entries of $\mathbf{M}$ are non-negative and $\mathbf{1}^T\mathbf{M} = \mathbf{1}^T$. Here $m_{ij}$ denotes the probability of transitioning from state $j$ to state $i$. Each column of $\mathbf{M}$ is a distribution: the $j^{th}$ column is the probability distribution over the states in the next step, given that the current state is $j$. If the starting state is $j$, the probability of being in state $i$ after $n$ steps is $m_{ij}^{(n)}$. If the current distribution over the states is $\mathbf{s}$, then after a transition, the distribution is transformed to $\mathbf{Ms}$. A distribution such that $\mathbf{Ms} = \mathbf{s}$ is called a \textit{stationary} distribution of the Markov Chain.

For any Markov Chain, we can construct a graph whose vertices correspond to the states, and there is an edge from vertex $i$ to $j$ if and only if $m_{ji} > 0$, i.e.\ there is a transition from state $i$ to state $j$ with nonzero probability. A Markov Chain is said to be \emph{ergodic} if its graph is aperiodic and irreducible (strongly connected), i.e.\ there exists an $N$ such that there is a path of length $N$ between any pair of (not necessarily distinct) vertices. An equivalent characterisation is: there exists an $N$ such that all the entries of $\mathbf{M}^N$ are strictly positive. For completeness, we note some spectral properties ergodic Markov Chains enjoy \cite[Chapter 4, Theorem 6]{gallagher}.
\begin{theorem}[Standard]
\label{spectral}
Let $\mathbf{M} \in \rationals^{k\times k}$ represent an ergodic Markov Chain. Then $\lambda_0 = 1$ is the largest real eigenvalue of $\mathbf{M}$, and $|\lambda| < 1$ for all other eigenvalues $\lambda$. Furthermore, $\lim_{n\rightarrow \infty}\mathbf{M}^n = \mathbf{S} = \mathbf{s}\mathbf{1}^T$, where $\mathbf{s}$ is the unique stationary distribution of $\mathbf{M}$.
\end{theorem}

\section{The Reduction}
In this section, we prove Theorem \ref{mainresult}. We shall start with the companion matrix $\mathbf{A}$ and the vector $\mathbf{u} = \begin{bmatrix} u_0 & \dots & u_{k-1}\end{bmatrix}^T$ of the initial values of the LRS, and construct the required ergodic Markov Chain. Recall from the preliminaries that $u_n = \mathbf{e_1}^T\mathbf{A}^n\mathbf{u}$. From the discussion in the preliminaries, we can also assume that the given LRS is non-degenerate with finitely many zeroes. In particular, by brute enumeration, we are guaranteed to find $k$ consecutive indices with nonzero entries. We will thus assume, without loss of generality, that $u_0, \dots, u_{k-1}$ are all nonzero.\footnote{This assumption can be made trivially if we are reducing the Skolem Problem. Further, for Positivity, we can trivially assume $u_0, \dots, u_{k-1} > 0$.}

The key idea is to construct a $(k+1)\times(k+1)$ stochastic matrix via a decomposition $\mathbf{M} = \mathbf{S} + \mathbf{D}$ (intuitively, Stationary distribution plus Disturbance) that represents an ergodic Markov Chain due to the following properties:
\begin{enumerate}
\item $\mathbf{S} =\mathbf{s1}^T$ is a stochastic matrix, each of whose columns are a distribution $\mathbf{s}$ with strictly positive entries. The choice of such $\mathbf{s}$ can be completely arbitrary: for simplicity, we choose each entry to be $1/(k+1)$.
\item $\mathbf{D}$ is a matrix such that $\mathbf{DS} = \mathbf{SD} = \mathbf{O}$, its entries are small enough to ensure that the all entries of $\mathbf{S} + \mathbf{D}$ are strictly positive, and finally, there exist indices $i, j$, and positive constants $\eta, \rho \in \rationals$ such that for all $n \ge 1$, $d_{ij}^{(n)} = \eta u_n/\rho^n$.
\end{enumerate}
From these properties, we first prove that $\mathbf{M}$ is indeed a stochastic matrix. Since all columns of $\mathbf{S}$ are identical and have strictly positive entries, the condition $\mathbf{SD} = \mathbf{O}$ is equivalent to $\mathbf{1}^T\mathbf{D} = \mathbf{0}^T$, i.e.\ all columns of $\mathbf{D}$ sum up to zero. We have that all entries of $\mathbf{M}$ are strictly positive, and $\mathbf{1}^T\mathbf{M} = \mathbf{1}^T(\mathbf{S}+\mathbf{D}) = \mathbf{1}^T$. Further, since every entry of $\mathbf{M}$ is strictly positive, it immediately follows that the Markov Chain is ergodic with $N=1$. 

We then observe that for all $n\ge 1$, $\mathbf{S}^n = \mathbf{S}$ and $\mathbf{M}^n = (\mathbf{S}+\mathbf{D})^n = \mathbf{S} + \mathbf{D}^n$, since $\mathbf{DS} = \mathbf{SD} = \mathbf{O}$. The property that $\mathbf{DS} = \mathbf{O}$ also gives us that $\mathbf{s}$ is the stationary distribution of $\mathbf{M}$: indeed, $\mathbf{MS} = \mathbf{S}^2 + \mathbf{DS} = \mathbf{S}$.  We let $r = s_{ij}$, and complete the reduction as follows (note that we actually reduce the complements of the Positivity and Ultimate Positivity Problems):
\begin{enumerate}
\item Skolem: Does there exist $n$ such that $m_{ij}^{(n)} = r$?
\item Positivity: Does there exist $n$ such that $m_{ij}^{(n)} < r$?
\item Ultimate Positivity: Do there exist infinitely many $n$ such that $m_{ij}^{(n)} < r$?
\end{enumerate}

Since $m_{ij}^{(n)} = s_{ij} + d_{ij}^{(n)} =  r + d_{ij}^{(n)}$, it only remains to construct $\mathbf{D}$ and choose indices $i, j$, and constants $\eta, \rho$. We do so in a top-down fashion. 

Let $\mathbf{C} \in \rationals^{(k+1)\times(k+1)}$ be a matrix, each of whose columns sum up to $0$. Similarly to the preceding discussion, it follows that $\mathbf{SC} = \mathbf{O}$. We argue that we can set $\mathbf{D}$ to be of the form $\frac{1}{\rho}(\mathbf{C} - \mathbf{CS})$, where $\rho \in \rationals$ is chosen large enough to ensure that all the entries of $\mathbf{S} + \mathbf{D}$ are strictly positive, e.g.\ if the smallest entry of $\mathbf{S}$ is $\sigma$ and the largest entry (by magnitude) of $\mathbf{C} - \mathbf{CS}$ is $\gamma$, then it suffices to choose $\rho = 2\gamma/\sigma$. It also immediately follows by construction that $\mathbf{SD} = \mathbf{O}$. Since we have noted that $\mathbf{S}^2 = \mathbf{S}$, we also see that $\mathbf{DS} = \mathbf{O}$. We can prove, by a simple induction, that for all $n \ge 1$, $$\rho^n\mathbf{D}^n = \mathbf{C}^n - \mathbf{C}^n \mathbf{S} = \mathbf{C}^n(\mathbf{I} - \mathbf{S}).$$
The base case is satisfied by construction. For the induction step, we express
\begin{align*}
\rho^{n+1}\mathbf{D}^{n+1} &= ( \mathbf{C}^n - \mathbf{C}^n \mathbf{S})(\mathbf{C} - \mathbf{CS}) \\
&= \mathbf{C}^{n+1} - \mathbf{C}^{n+1}\mathbf{S} - \mathbf{C}^{n}\mathbf{SC} +  \mathbf{C}^{n}\mathbf{SCS} \\
&= \mathbf{C}^{n+1} - \mathbf{C}^{n+1}\mathbf{S}
\end{align*}
using the fact that $\mathbf{SC} = \mathbf{O}$. Now, to ensure that $d_{ij}^{(n)} = \eta u_n/\rho^n$ for all $n \ge 1$, we can equivalently choose $\mathbf{C}, \eta, i, j$ such that $\mathbf{e_i}^T\mathbf{C}^n(\mathbf{I} - \mathbf{S})\mathbf{e_j} = \eta u_n = \mathbf{e_1}^T\mathbf{A}^n(\eta\mathbf{u})$.

Although $\mathbf{C} \in \rationals^{(k+1)\times(k+1)}$ and $\mathbf{A} \in \rationals^{k\times k}$ have different dimensions, we begin to see a correspondence. We shall pick $i =1$ and ``map'' $\eta \mathbf{u}$ to $(\mathbf{I} - \mathbf{S})\mathbf{e_j}$. Since the former is a $k$ dimensional vector, the intuition is to only consider the first $k$ entries of $(\mathbf{I} - \mathbf{S})\mathbf{e_j}$. Thus, we shall choose $\mathbf{C}$ to be a matrix of the form 
$$
\begin{bmatrix}
\mathbf{B} & \mathbf{0} \\
-\mathbf{1}^T\mathbf{B} & 0 
\end{bmatrix}
$$
where $\mathbf{B} \in \rationals^{k\times k}$. It is easy to see that $\mathbf{1}^T\mathbf{C} = \mathbf{0}^T$ for such a choice. We note, by a simple induction, that for $n \ge 1$
$$
\mathbf{C}^n = 
\begin{bmatrix}
\mathbf{B}^n & \mathbf{0} \\
-\mathbf{1}^T\mathbf{B}^n & 0 
\end{bmatrix}.
$$
The base case is satisfied by construction. For the inductive step, simplify
\begin{align*}
\mathbf{C}^{n+1} &= 
\begin{bmatrix}
\mathbf{B}^n & \mathbf{0} \\
-\mathbf{1}^T\mathbf{B}^n & 0 
\end{bmatrix}
\begin{bmatrix}
\mathbf{B} & \mathbf{0} \\
-\mathbf{1}^T\mathbf{B} & 0 
\end{bmatrix} \\
&=
\begin{bmatrix}
\mathbf{B}^{n}\cdot \mathbf{B} +\mathbf{0}\cdot(-\mathbf{1}^T\mathbf{B}) & \mathbf{B}^n\cdot\mathbf{0} + \mathbf{0}\cdot 0\\
(-\mathbf{1}^T\mathbf{B}^{n})\cdot\mathbf{B} + 0 \cdot (-\mathbf{1}^T\mathbf{B}) & (-\mathbf{1}^T\mathbf{B}^n)\cdot \mathbf{0} + 0 \cdot 0 
\end{bmatrix} \\
&=
\begin{bmatrix}
\mathbf{B}^{n+1} & \mathbf{0} \\
-\mathbf{1}^T\mathbf{B}^{n+1} & 0 
\end{bmatrix}.
\end{align*}

Let $\mathbf{y_j} \in \rationals^{k}$ denote the vector obtained by deleting the last entry of $(\mathbf{I}- \mathbf{S})\mathbf{e_j}$. In other words, $\mathbf{y_j}$ is the vector formed by the first $k$ entries of the $j^{th}$ column of $\mathbf{I} - \mathbf{S}$. We must now have that $\mathbf{e_1}^T\mathbf{A}^n(\eta\mathbf{u}) = \mathbf{e_1}^T\mathbf{C}^n(\mathbf{I}- \mathbf{S})\mathbf{e_j} = \mathbf{e_1}^T\mathbf{B}^n\mathbf{y_j}$, where the choice is over $\mathbf{B} \in \rationals^{k\times k}$. 

It suffices to pick $\mathbf{B}$ of the form $\mathbf{F}^{-1}\mathbf{AF}$: this gives $\mathbf{B}^n = \mathbf{F}^{-1}\mathbf{A}^n\mathbf{F}$. We then require that for all $n \ge 1$, we must have 
$$
\mathbf{e_1}^T\mathbf{A}^n(\eta \mathbf{u}) = (\mathbf{e_1}^T\mathbf{F}^{-1})\mathbf{A}^n(\mathbf{Fy_j}).
$$
We shall choose $\mathbf{F}, \eta > 0, j$ such that $\mathbf{F}$ is a diagonal matrix, $\mathbf{e_1}^T\mathbf{F}^{-1} = \mathbf{e_1}^T$ ($f_{11} = 1$ ensures this), and $\mathbf{Fy_j} = \eta\mathbf{u}$. Recall that we assume all the entries of $\mathbf{u}$ are nonzero, and the entries of $\mathbf{y_j}$ are $-s_1, \dots, 1-s_j, \dots, -s_k$, which are also nonzero courtesy our choice of $\mathbf{S}$. Since $f_{11} = 1$ and $\eta$ must be positive, $u_0$ and the first entry of $\mathbf{y_j}$ must have the same sign. If $u_0 > 0$, we choose $j=1$, and $\eta = (1-s_1)/u_0$. Otherwise, $u_0 < 0$, in which case we choose $j=2$, and $\eta = -s_1/u_0$. Let the entries of $\mathbf{y_j}$ thus chosen be $\mu_1, \dots, \mu_k$. Note that $\eta u_0/\mu_1 = 1$. Taking $\mathbf{F} = \diag(1, \eta u_1/\mu_2, \dots, \eta u_{k-1}/\mu_k)$, and thus $\mathbf{F}^{-1} = \diag(1, \mu_2/(\eta u_1), \dots, \mu_k/(\eta u_{k-1}))$ ensures that $\mathbf{e_1}^T\mathbf{F}^{-1} = \mathbf{e_1}^T$ and $\mathbf{Fy_k} = \eta\mathbf{u}$. Propagating these choices to construct $\mathbf{B}, \mathbf{C}$, identify $\rho$, construct $\mathbf{D}$ and finally $\mathbf{M}$ completes our reduction.

\section{Discussion}
To put our result in perspective, it is worth noting that Markov Reachability Problems are a special class of model checking problems for Linear Dynamical Systems as described in \cite{linearloops}. The problem considers the orbit $\seq{\mathbf{M}^n\mathbf{s}}_{n=0}^\infty$ given by a matrix and a vector, a finite set of regions given by polynomial inequalities, and asks whether the characteristic $\omega$-word $\alpha$ that records which of the regions the orbit visits at every time step belongs to a given $\omega$-regular language. In our setting, $\mathbf{M}$ is stochastic, and the regions, which are affine hyperplanes or halfspaces, cannot be handled by the techniques of \cite{linearloops}. Techniques have been developed to solve these problems for diagonalisable $\mathbf{M}$ and prefix-independent $\omega$-regular languages \cite{prefixindependent}. We assume neither prefix independence nor diagonalisability. The problem for diagonalisable $\mathbf{M}$ and general $\omega$-regular languages is shown to reduce to the Positivity Problem for simple LRS \cite{powerpositivity}, which is analogous but incomparable to the result we show.

\bibliographystyle{elsarticle-num}
\bibliography{MarkovChains}

\begin{thebibliography}{10}
\expandafter\ifx\csname url\endcsname\relax
  \def\url#1{\texttt{#1}}\fi
\expandafter\ifx\csname urlprefix\endcsname\relax\def\urlprefix{URL }\fi
\expandafter\ifx\csname href\endcsname\relax
  \def\href#1#2{#2} \def\path#1{#1}\fi

\bibitem{Baier-Katoen}
C.~Baier, J.-P. Katoen, {Principles of model checking}, MIT Press, 2008.

\bibitem{mdp-rl}
L.~P. Kaelbling, M.~L. Littman, A.~W. Moore, {Reinforcement learning: a
  survey}, J. Artif. Int. Res. 4~(1) (1996) 237–285.

\bibitem{Agrawal2015}
M.~Agrawal, S.~Akshay, B.~Genest, P.~S. Thiagarajan, {Approximate Verification
  of the Symbolic Dynamics of Markov Chains}, Journal of the ACM (JACM) 62
  (2015) 1 -- 34.
\newblock \href {https://doi.org/10.1145/2629417} {\path{doi:10.1145/2629417}}.

\bibitem{logicprob}
D.~Beauquier, A.~Rabinovich, A.~Slissenko, {A Logic of Probability with
  Decidable Model Checking}, J. Log. Comput. 16 (2006) 461--487.
\newblock \href {https://doi.org/10.1093/logcom/exl004}
  {\path{doi:10.1093/logcom/exl004}}.

\bibitem{KVAK10}
V.~A. Korthikanti, M.~Viswanathan, G.~Agha, Y.-M. Kwon, {Reasoning about MDPs
  as Transformers of Probability Distributions}, in: QEST, 2010, pp. 199--208.
\newblock \href {https://doi.org/10.1109/QEST.2010.35}
  {\path{doi:10.1109/QEST.2010.35}}.

\bibitem{KwonGul}
Y.~Kwon, G.~Agha, {Linear Inequality LTL (iLTL): A Model Checker for Discrete
  Time Markov Chains}, in: J.~Davies, W.~Schulte, M.~Barnett (Eds.), Formal
  Methods and Software Engineering, Springer Berlin Heidelberg, 2004, pp.
  194--208.
\newblock \href {https://doi.org/10.1007/978-3-540-30482-1\_21}
  {\path{doi:10.1007/978-3-540-30482-1\_21}}.

\bibitem{joeljames3}
J.~Ouaknine, J.~Worrell, {Positivity Problems for Low-Order Linear Recurrence
  Sequences}, in: Proceedings of the Twenty-Fifth Annual {ACM-SIAM} Symposium
  on Discrete Algorithms, {SODA} 2014, {SIAM}, 2014, pp. 366--379.
\newblock \href {https://doi.org/10.1137/1.9781611973402.27}
  {\path{doi:10.1137/1.9781611973402.27}}.

\bibitem{skolem}
M.~Mignotte, T.~N. Shorey, R.~Tijdeman, {The distance between terms of an
  algebraic recurrence sequence}, Journal f{\"u}r die reine und angewandte
  Mathematik 349 (1984) 63--76.

\bibitem{skolemv}
N.~K. Vereshchagin, {The problem of appearance of a zero in a linear recurrence
  sequence }, Matematicheskie Zametki (in Russian) 38 (1985) 177--189.
\newblock \href {https://doi.org/10.1007/BF01156238}
  {\path{doi:10.1007/BF01156238}}.

\bibitem{bilu2022skolem}
Y.~Bilu, F.~Luca, J.~Nieuwveld, J.~Ouaknine, D.~Purser, J.~Worrell, {Skolem
  Meets Schanuel}, in: 47th International Symposium on Mathematical Foundations
  of Computer Science (MFCS 2022), Vol. 241 of LIPIcs, Schloss Dagstuhl --
  Leibniz-Zentrum f{\"u}r Informatik, 2022, pp. 20:1--20:15.
\newblock \href {https://doi.org/10.4230/LIPIcs.MFCS.2022.20}
  {\path{doi:10.4230/LIPIcs.MFCS.2022.20}}.

\bibitem{ouaknine2014ultimate}
J.~Ouaknine, J.~Worrell, {Ultimate Positivity is Decidable for Simple Linear
  Recurrence Sequences}, in: Automata, Languages, and Programming - 41st
  International Colloquium, {ICALP} 2014, Proceedings, Part {II}, Vol. 8573 of
  LNCS, Springer, 2014, pp. 330--341.
\newblock \href {https://doi.org/10.1007/978-3-662-43951-7\_28}
  {\path{doi:10.1007/978-3-662-43951-7\_28}}.

\bibitem{ouaknine2014positivity}
J.~Ouaknine, J.~Worrell, {On the positivity problem for simple linear
  recurrence sequences}, in: International Colloquium on Automata, Languages,
  and Programming, Springer, 2014, pp. 318--329.
\newblock \href {https://doi.org/10.1007/978-3-662-43951-7\_27}
  {\path{doi:10.1007/978-3-662-43951-7\_27}}.

\bibitem{turakainen}
P.~Turakainen,
  \href{https://api.semanticscholar.org/CorpusID:13136683}{{Generalized
  automata and stochastic languages}}, in: Proceedings of the American
  Mathematical Society, Vol.~21, 1969, pp. 303--309.
\newline\urlprefix\url{https://api.semanticscholar.org/CorpusID:13136683}

\bibitem{oldresult}
S.~Akshay, T.~Antonopoulos, J.~Ouaknine, J.~Worrell, {Reachability problems for
  Markov chains}, Information Processing Letters 115~(2) (2015) 155--158.
\newblock \href {https://doi.org/https://doi.org/10.1016/j.ipl.2014.08.013}
  {\path{doi:https://doi.org/10.1016/j.ipl.2014.08.013}}.

\bibitem{akshaycav2023}
S.~Akshay, K.~Chatterjee, T.~Meggendorfer, D.~Zikelic, {MDPs as Distribution
  Transformers: Affine Invariant Synthesis for Safety Objectives}, in: Computer
  Aided Verification: 35th International Conference, CAV 2023, Proceedings,
  Part III, Springer-Verlag, Berlin, Heidelberg, 2023, p. 86–112.
\newblock \href {https://doi.org/10.1007/978-3-031-37709-9\_5}
  {\path{doi:10.1007/978-3-031-37709-9\_5}}.

\bibitem{piribauer}
J.~Piribauer, C.~Baier, {On Skolem-Hardness and Saturation Points in Markov
  Decision Processes}, in: 47th International Colloquium on Automata,
  Languages, and Programming (ICALP 2020), Vol. 168 of LIPIcs, Schloss Dagstuhl
  -- Leibniz-Zentrum f{\"u}r Informatik, Dagstuhl, Germany, 2020, pp.
  138:1--138:17.
\newblock \href {https://doi.org/10.4230/LIPIcs.ICALP.2020.138}
  {\path{doi:10.4230/LIPIcs.ICALP.2020.138}}.

\bibitem{prefixindependent}
S.~Almagor, T.~Karimov, E.~Kelmendi, J.~Ouaknine, J.~Worrell, {Deciding
  $\omega$-Regular Properties on Linear Recurrence Sequences}, Proc. ACM
  Program. Lang. 5~(POPL) (Jan 2021).
\newblock \href {https://doi.org/10.1145/3434329} {\path{doi:10.1145/3434329}}.

\bibitem{powerpositivity}
T.~Karimov, E.~Kelmendi, J.~Nieuwveld, J.~Ouaknine, J.~Worrell, {The Power of
  Positivity}, in: 2023 38th Annual ACM/IEEE Symposium on Logic in Computer
  Science (LICS), 2023, pp. 1--11.
\newblock \href {https://doi.org/10.1109/LICS56636.2023.10175758}
  {\path{doi:10.1109/LICS56636.2023.10175758}}.

\bibitem{skolem-sml}
T.~Skolem, {Einige S{\"a}tze {\"u}ber $\pi$-adische Potenzreihen mit Anwendung
  auf gewisse exponentielle Gleichungen}, Mathematische Annalen 111~(1) (1935)
  399--424.
\newblock \href {https://doi.org/10.1007/BF01472228}
  {\path{doi:10.1007/BF01472228}}.

\bibitem{mahler-sml}
K.~Mahler, {An arithmetic property of Taylor coefficients of rational functions
  (1935)}, in: M.~Baake, Y.~Bugeaud, M.~Coons (Eds.), The Legacy of Kurt
  Mahler, {EMS} Press, 2019, pp. 437--448.
\newblock \href {https://doi.org/10.4171/DMS/8} {\path{doi:10.4171/DMS/8}}.

\bibitem{lech-sml}
C.~Lech, {A note on recurring series}, Arkiv f{\"o}r Matematik 2~(5) (1953)
  417--421.
\newblock \href {https://doi.org/10.1007/BF02590997}
  {\path{doi:10.1007/BF02590997}}.

\bibitem{berstelmignotte}
J.~Berstel, M.~Mignotte, {Deux propri{\'e}t{\'e}s d{\'e}cidables Des Suites
  r{\'e}currentes lin{\'e}aires}, Bulletin de la Societe mathematique de France
  79 (1976) 175--184.
\newblock \href {https://doi.org/10.24033/bsmf.1823}
  {\path{doi:10.24033/bsmf.1823}}.

\bibitem{everest}
G.~Everest, A.~{van der Poorten}, I.~Shparlinski, T.~Ward, {Recurrence
  sequences}, Vol. 104 of Mathematical Surveys and Monographs, American
  Mathematical Society, United States, 2003.

\bibitem{gallagher}
R.~G. Gallager, {Discrete Stochastic Processes}, Springer New York, NY, 1996.
\newblock \href {https://doi.org/https://doi.org/10.1007/978-1-4615-2329-1}
  {\path{doi:https://doi.org/10.1007/978-1-4615-2329-1}}.

\bibitem{linearloops}
T.~Karimov, E.~Lefaucheux, J.~Ouaknine, D.~Purser, A.~Varonka, M.~A. Whiteland,
  J.~Worrell, {What’s Decidable about Linear Loops?}, Proc. ACM Program.
  Lang. 6~(POPL) (Jan 2022).
\newblock \href {https://doi.org/10.1145/3498727} {\path{doi:10.1145/3498727}}.

\end{thebibliography}
\end{document}